%% file: ms.tex
\documentclass[sigconf]{acmart}
\pdfoutput=1
\usepackage{booktabs} 

\PassOptionsToPackage{hyphens}{url}

\newcommand{\xcomment}[1]{{}}
\newcommand{\para}[1]{\smallskip\noindent {\bf #1} }

\setcopyright{rightsretained}

\copyrightyear{2018} 
\acmYear{2018} 
\setcopyright{acmcopyright}
\acmConference[ICSE-NIER'18]{40th International Conference on Software Engineering: New Ideas and Emerging Results Track}{May 27-June 3}{Gothenburg, Sweden}
\acmBooktitle{ICSE-NIER'18: 40th International Conference on Software Engineering: New Ideas and Emerging Results Track, May 27-June 3, 2018, Gothenburg, Sweden}

\begin{document}
\title{Mining Container Image Repositories for Software Configuration and Beyond}
\titlenote{This is an extended version of the paper presented at ICSE-NIER '18.}

\author{Tianyin Xu and Darko Marinov}
\affiliation{%
  \institution{University of Illinois at Urbana-Champaign}
}
\email{{tyxu, marinov}@illinois.edu}

\renewcommand{\shortauthors}{T. Xu and D. Marinov}

\begin{abstract}
This paper introduces the idea of mining container image repositories for configuration and other
deployment information of software systems.
Unlike traditional software repositories (e.g., source code repositories and app stores), 
image repositories encapsulate the entire execution ecosystem for 
running target software, including its configurations, dependent libraries and components, 
and OS-level utilities, which contributes to a wealth of data and information.  
We showcase the opportunities based on concrete software engineering tasks that 
can benefit from mining image repositories.
To facilitate future mining efforts, we summarize the challenges of analyzing image repositories
and the approaches that can address these challenges.
We hope that this paper will stimulate exciting
research agenda of mining this emerging type of software repositories.
\end{abstract}

%
%



\maketitle

\input{intro}
\input{background}

\input{usecase}
\input{challenge}

\input{method}
\input{limitation}

\input{related}
\input{conclusion}

\bibliographystyle{acm}
\bibliography{ref} 

\end{document}

%% file: intro.tex
\section{Introduction}
\label{sec:intro}

Mining software repositories (MSR) has been proven to be an effective approach for discovering, characterizing,
and understanding software engineering practices, towards improving software productivity and quality.
Existing MSR studies mostly focus on {\it software development}
by mining code repositories (including source code, commit histories, and bug reports)~\cite{xie:09,hassan:08,nachi:09}
and {\it software release} by mining app stores and package repositories~\cite{harman:12,abate:15}.
Few studies cover field configurations of software systems (e.g., for deployment and orchestration).
In fact, information of field configurations is highly desired, not only by operators and sysadmins to learn best practices, 
but also by developers and DevOps engineers to measure software usability and manageability~\cite{parnin:17,xu:15:2,xu:17,xu:13,hilton:17,raab:17}.

\begin{table}[t] \small
\centering
\caption{A comparison of image versus code repositories.}
\vspace{-5pt}
\begin{tabular}{l|l|l} 
  \toprule
 & {\bf Container image repo} & {\bf Source code repo} \\
\midrule
Usage               &  system operation                & software development  \\
User                &  sysadmins and operators         & developers            \\
Store               &  Docker Hub, Docker Store, ...   & GitHub, Sourceforge, ...   \\
Content             &  executables + exec. context     & source code           \\
Configuration       &  customized                      & default/customizable  \\
Scope               &  entire software stack           & specific project      \\
Evolution           &  different software versions     & source code changes   \\
\bottomrule
\end{tabular}  
\label{tab:comparison}
\end{table}

One fundamental obstacle to the study of configurations lies in the fact that traditional software repositories such as source code repositories and
app stores contain little information of how the software is actually being used in the wild. 
Historically, studying field configurations used ethnographic methods~\cite{haber:07,barrett:04} and manual
data collection from second-hand data sources~\cite{xu:15:2,xu:17}.
For example, a study of how software is configured in the field~\cite{xu:15:2}
took six person-month to collect configuration files attached in issue reports on mailing lists and online forums.
However, this dataset, despite the only one of its kind, is highly biased to misconfiguration cases 
and is incomplete---it is hard to determine the values referencing to execution context (e.g., environment variables and file content).


In this paper, we advocate that container image repositories, as an emerging type of software repositories, provide a plethora of opportunities
to study configurations and other field operations for a variety of software.
Unlike source code repositories for software development, container images
are used for operations. 
A container image is defined as a stand-alone, executable package of a piece of software\footnote{Containers
are often designed for the microservice architecture in which each container runs one software service, so each image has its target software.} 
that includes everything needed to run it: binary code, configuration files, system libraries, language runtime, 
and management tools. Most of this information is not directly included in traditional software repositories.
Table~\ref{tab:comparison} compares container image repositories with traditional source code repositories.

Most importantly, the wide adoption of containerization techniques
drives the proliferation of image sharing.
According to Docker Hub's statistics, 
it has hosted 100K+ 
public image repositories contributing to 900K+ images, serving 12+ billion image pulls per week. 
Besides a small number of {\it official} image repositories
from certified software vendors (e.g., Apache, Oracle, and Red Hat), 
most of the repositories are shared by  
individual users and organizations, containing various customization, integration, and orchestration, to serve their own use cases.
These images are supposed to be directly invoked to create running containers,
without the need to compile or configure the software---``{\it building, shipping, and running any apps, anywhere}~\cite{docker}.''
Therefore, these image repositories form a massive information base of configuration and operation practices for mining and analysis.

We present the {\it opportunities}, {\it challenges}, and {\it methods} 
for mining image repositories based on our experience of working with image data.
We focus on repositories of Docker-based container images (a.k.a., {\it Docker images}), 
the {\it de facto} image format adopted in industry, and Docker Hub as the 
current largest online registry service for {\it public} Docker images\footnote{There are other online image registries such as Docker Store, Google Container Registry, and AWS Container Registry.}.
Our objective is to showcase the rich data and information encoded in image
repositories, and more importantly,
describe how several software engineering tasks---ranging from configuration design to dependency modeling to software orchestration to combinatorial testing---can potentially benefit from or be enabled by mining these repositories (cf. \S\ref{sec:usecases}).\footnote{Our initial focus is on understanding software configurations, 
and we plan to use the image mining infrastructure to address other software engineering tasks.}
To facilitate future mining efforts, we summarize the challenges of
mining image repositories (cf. \S\ref{sec:challenge}) and the methods that can address these challenges (cf. \S\ref{sec:collection}).
We hope that this paper will stimulate exciting research agenda of mining the emerging image repositories.



%% file: background.tex
\section{Image Repositories}
\label{sec:background}

This section goes over several preliminaries of container images and their repositories from the perspective of
mining and analysis, which establishes the context necessary to understand
the technical content presented in the subsequent sections.

\begin{figure}[t]
\centering
  \includegraphics[width=0.47\textwidth]{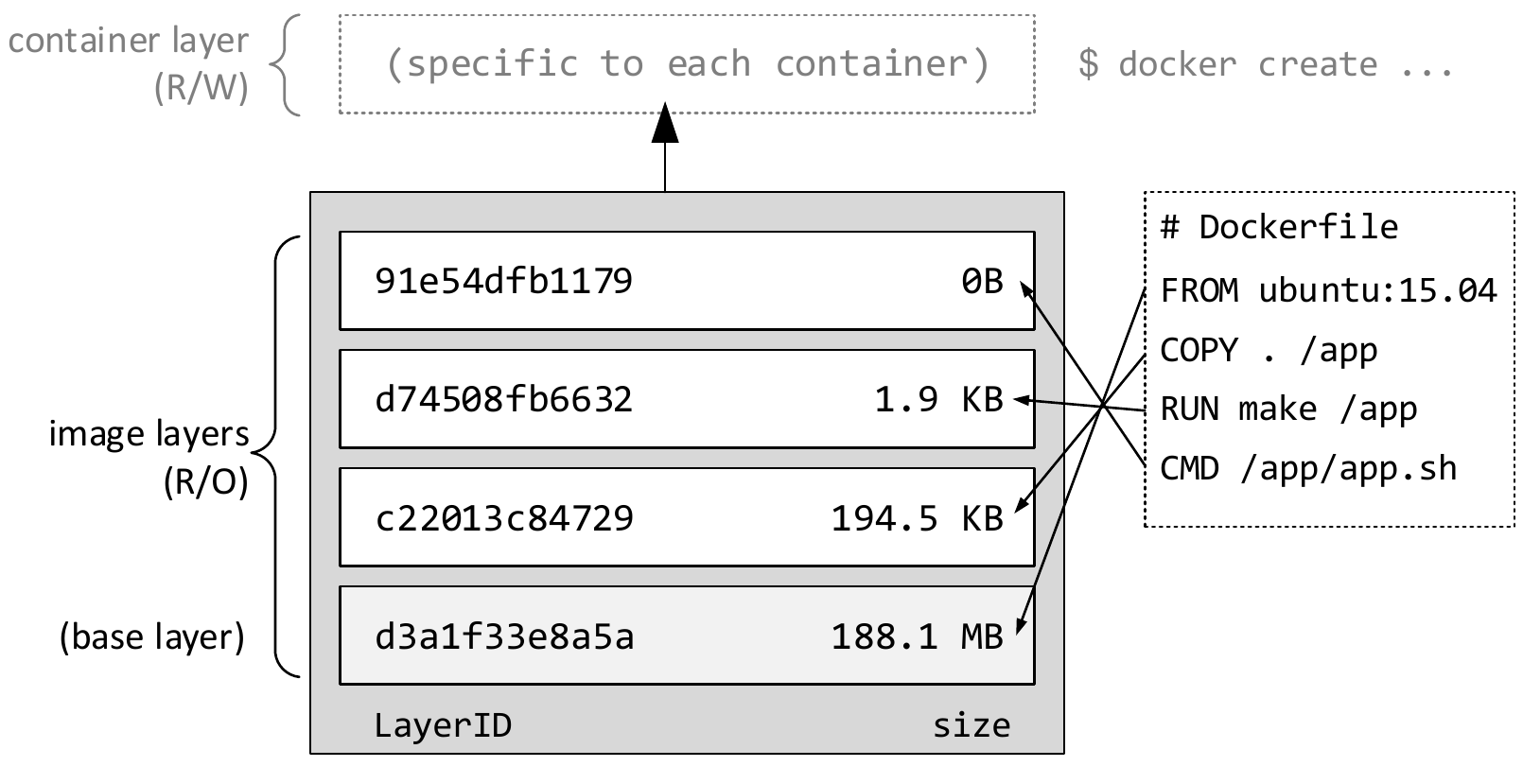}
  \caption{\small {\bf An example of a Docker image consisting of multiple layers corresponding to the Dockerfile instructions.}}
  \label{fig:image_layers}
\end{figure}

\para{Image organization.} In essence, an image is a filesystem-level snapshot that includes all the files needed for launching a running 
system instance (i.e., a container).
For Docker, images are organized as a series of {\it layers} stacking on top of one another.
Each layer is created by a build instruction specified in the image's Dockerfile (i.e., Docker's build file for specifying the instructions that can be executed to assemble an image, similar to Makefile for building an executable from source code). Each layer consists of the filesystem \texttt{diff} (files added or deleted) introduced by executing that instruction on the layer below it.
Stacking all the layers comprises the unified view of the image.
Note that layers (identified by unique LayerIDs) 
can be shared across multiple images, 
e.g., one can create a new image by adding new files onto the \texttt{ubuntu} image, and the new image shares all the layers of \texttt{ubuntu}. 
Figure~\ref{fig:image_layers} illustrates how an image is constructed through layers, with an example 
from Docker's official documentation~\cite{cow}.
Each instruction in the Dockerfile creates a layer, starting with the base layer as the \texttt{ubuntu} image.
All the layers inside the image are read-only. When a container is launched from the image, 
a read-write layer will be created on top of the image layers (which is specific to the container). 

An image can be \texttt{pull}ed from and \texttt{push}ed to registry services such as Docker Hub.
The image's metadata (e.g., version, LayerIDs, size, update time) and Dockerfile can be fetched through the \texttt{inspect} command or
the Docker's REST APIs.

\para{Image repositories.} An image repository on Docker Hub contains multiple images with different {\it tags} 
(typically used for annotating versions). An image on Docker Hub is identified by the repository name and the tag,
for example, \texttt{ubuntu:16.04} refers to the image with the tag \texttt{16.04} in the \texttt{ubuntu} repository.
All the tags, together with other metadata of the repositories 
(e.g., description, maintainer, community rating and comments, and update time) 
can be queried through the Docker's REST APIs.

Image repositories can be searched based on keywords.
Although Docker Hub does not provide the entire list of image repositories, Shu et al.~\cite{shu:16}
show that a dictionary-based search method can collect the vast majority of public repositories on Docker Hub.

\para{Containers.} Containers are runtime instances launched by \texttt{docker run} images 
(with parameters specifying network settings, restart policies, 
resource constraints, security settings, etc).
A container's flat filesystem differs from the original image which is
organized in layers. 
Moreover, the container creates (virtual) files for device drivers and \texttt{procfs} (\texttt{/dev} and \texttt{/proc}) based on 
the host OS, which are not included in images.
Also, containers typically execute initial instruction (specified in Dockerfiles)
to run the target software, which creates new files (e.g., logs and traces).

%% file: usecase.tex
\section{Opportunities}
\label{sec:usecases}

In this section, we showcase 
the research opportunities for software engineering enabled by the unique data encoded in container image repositories. 
We start from our initial focus---understanding software configurations by mining image repositories,
and envision such mining efforts to go beyond and be broadly applied to other software engineering tasks, 
including (but not limited to) dependency modeling, software orchestration, and combinatorial testing.
Note that container images are supposed to run out of the box, without the need of additional configuration
efforts---the configurations in image repositories are working samples rather than demos.

\para{Creating a feedback loop for configuration design.}
One key aspect of configuration design is the trade-off
between flexibility (configurability) and complexity (usability), which should be carefully made
with a user-centric design philosophy, as configuration is essentially an interface
for users to control and customize software behavior~\cite{xu:15:2,xu2:16}.
Feedback loops should be created to help developers understand how their software is 
configured in the wild, in order to tune the usability accordingly.
In addition, one can learn design lessons by comparatively studying the configuration characteristics 
of multiple software systems, and verify design hypothesis by selectively studying configurations of interests.

Furthermore, as shown in prior work~\cite{sai:14}, configuration requirements
can change over time---a correct configuration value in an old version 
could be obsolete or become invalid (producing undesired behavior) after
software upgrade. Understanding the characteristics of configuration changes through software evolution
is critical to software configuration design and maintenance. 

Historically, attacking the above problems is difficult,
especially for open-source software projects, due to
the lack of publicly available datasets~\cite{xu:15}. 
Unlike source code for which there are many open-source online repositories (e.g., GitHub),
software configurations are independently maintained by sysadmins and operators
who have no incentives to share their settings. Some companies do collect customers'
configuration values, but few of them are willing to open such data to public as
configuration settings often contain sensitive, confidential information.
Therefore, existing studies on field configurations are either by the companies (which are specific to one or two products), or based on tedious, time-consuming data collection effort (as discussed in \S\ref{sec:intro}).

With container image repositories, the usage statistics of configuration parameters can be collected by analyzing the configuration files
in the image repositories built for the same piece of software.
For popular software (e.g., those studied in~\cite{xu:15:2}), 
there are typically thousands of image repositories
made for different use cases
and scenarios, containing a diverse set of configuration settings.\footnote{As a comparison, \texttt{mysqld} and \texttt{httpd} studied in~\cite{xu:15:2} have 9133 and 2006 image repositories (which contain the corresponding configuration files with different versions of the software) on Docker Hub, respectively, while the dataset in ~\cite{xu:15:2} only contains 823 and 311 configuration files of these two software projects, respectively.}  
Moreover, as image repositories contain different versions of the target software and 
the configurations working for each version, mining these repositories enables the opportunities to 
understand software configuration with software evolution in depth. 

\para{Modeling cross-component configuration dependencies.} Misconfigurations across multiple software stacks or components
    are among the most urgent but thorny problems in software reliability~\cite{sayagh:17,xu:15}. One fundamental obstacle in dealing
    with these misconfigurations lies in the challenges of understanding and modeling dependencies of configurations
    across components. Existing studies attempt to understand cross-component dependencies based on user-reported issues posted on mailing lists and online forums~\cite{sayagh:17}. However, the user-generated data cannot help understand the unknown unknowns or model the complete dependency information, 
not to mention the tremendous overhead of collect them. 
    
Mining image repositories provides opportunities of unraveling such information, 
as images encapsulate the complete environment for running target software from the OS kernel to user-level applications.
Many images are built for system infrastructure made up of different components (each
as a microservice) that have been configured to work with each other.
Therefore, image repositories provide an open dataset of rich, extensive, and concrete configuration values recorded
in configuration files, databases, and system environment.
More importantly, unlike a second-hand dataset in which configuration values are treated as 
isolated string literals, image repositories associate these values with their context, including
the executable code, resources/entities referenced by these values, and dependent software components.


\para{Discovering software orchestration.} 
Unlike source code repositories dedicated for a specific piece of software, 
image repositories often serve as building blocks for large-scale, complex systems composed of multiple software components.
These software components can either be packed into a single image (e.g., the image \texttt{wordpress:php7.1-apache} as a web stack),
or form distributed systems running on top of multiple images maintained in separate repositories (e.g., the Hadoop-based data processing framework published
by \texttt{uhopper} that is composed by 
\texttt{hadoop-namenode}, \texttt{hadoop-datanode}, \texttt{hadoop-spark}, and
other \texttt{hadoop-*} repositories).
Therefore, image repositories are great resources for studying how different software components (and their versions)
are glued together and orchestrated as a service. 
Such study can not only reveal the field practices of glue logic planned by software developers,
but also potentially discover spontaneous use cases invented by power users. 




Note that for the case of multiple images, it takes additional effort to collect orchestration information of these images,
as each image by itself does not explicitly specify the other images it connects to. One data source are ``compose files''
used by \texttt{docker} \texttt{compose} which specify how multiple containers are orchestrated from images.

\para{Improving combinatorial testing and tuning.} 
Mining image repositories can be used to understand common combinations and value distributions of 
binaries and configurations, in order to help test prioritization, performance tuning, and/or security auditing.

Testing of configurable software (e.g., a software product line, SPL) requires 
not only executing the software for certain inputs but also applying these inputs with various combinations of features. 
One key challenge is to select the subset of combinations that are representative and cover typical use cases,
as testing all possible combinations is not feasible (e.g., an SPL with 10 configurable features can have more than $2^{10}$ distinct configurations).
While combinatorial methods can explore various combinations of configurations, 
they are still quite costly~\cite{kim:13,dumlu:11,henard:14,halin:17}, and may focus on irrelevant combinations rarely used in practice. 
Mining image repositories can discover combinations that are actually used, 
allowing both speeding up testing and finding bugs for relevant configurations.

While combinatorial testing for functional correctness requires checking all combinations that arise in practice, 
performance tuning can be biased toward the most frequent configuration settings to optimize expected runtime 
(over the distribution of configurations).
Understanding how the software is actually used can also help developers better tune the performance
of the software by focusing on common systems environment and configuration settings.

The similar idea can be applied to security auditing---if the content of an executable file differs from
all files with the same name or path in the vast pool of image repositories, it is suspicious.

\para{Using images as test beds for software engineering tools.} Image repositories can serve as real-world test beds for research tools,
including misconfiguration detection, binary analysis for malware detection, portability testing, performance auto-tuning, etc. 
Taking misconfiguration detection
as an example, existing research efforts mostly evaluate the proposed methods and tools on self-injected
errors or a small set of known misconfigurations~\cite{xu:16,zhang:14,santo:16,santo:17}.
However, it is hard to measure the actual benefits in large-scale real-world deployments.
Image repositories can be used to quantitatively answer such questions, as they
themselves form a diverse and comprehensive dataset of real-world configurations and their context. 
We envision such test beds to be built on top of existing container image repositories.

\xcomment{
\para{Many more!} The above cases are just three examples of software engineering tasks
that can benefit from mining image repositories. 
On the other hand, the opportunities are by no means limited to these---many other research problems
related to software's field operations can potentially be studied through image repositories. 
}

%% file: challenge.tex

\xcomment{
\begin{figure*}[t]
\centering
\begin{minipage}[t]{0.315\textwidth}
	\centering   
  \includegraphics[width=0.99\textwidth]{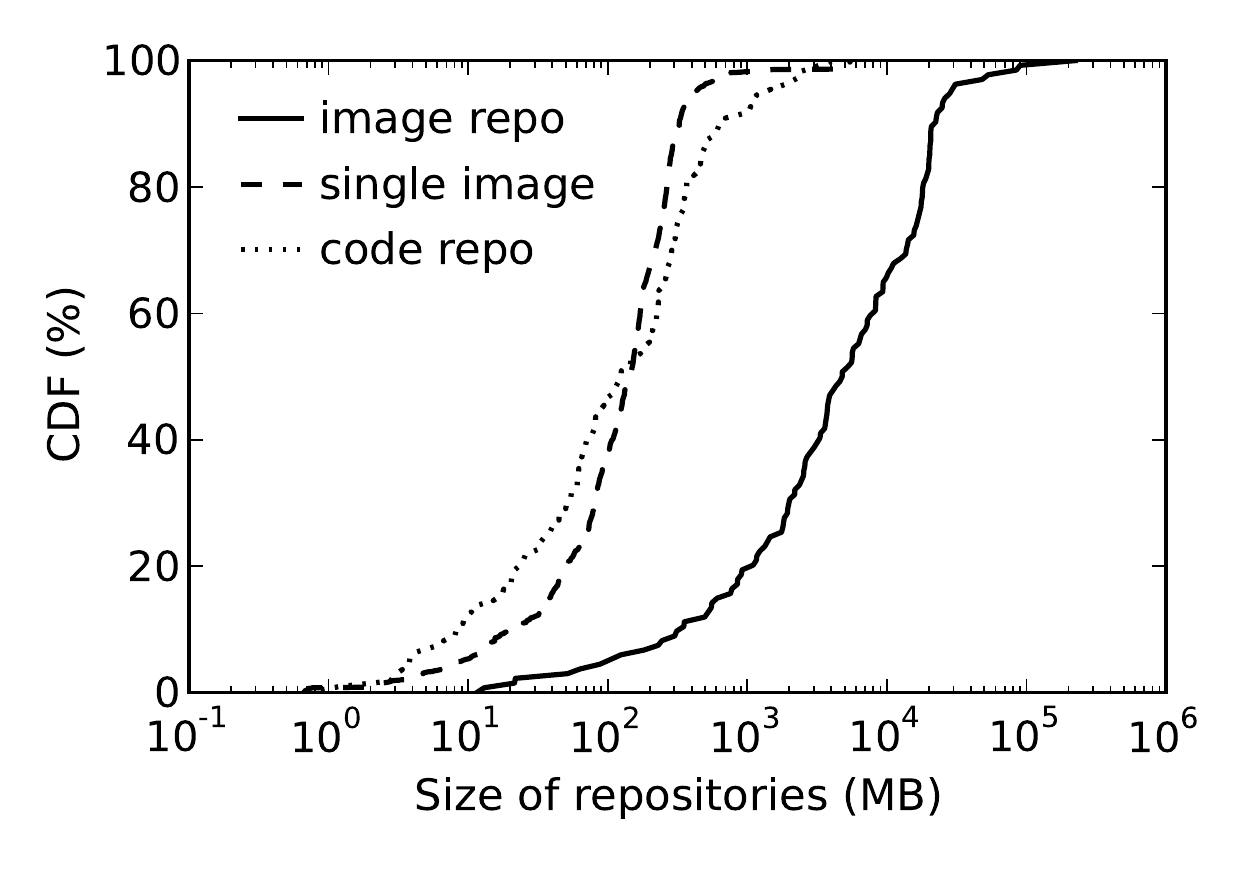}
  \vspace{-20pt}
  \caption{\small {\bf Size of image repositories versus\\ code repositories.}} 
  \label{fig:repo_size}
\end{minipage}
\hfill
\begin{minipage}[t]{0.315\textwidth}
	\centering   
  \includegraphics[width=0.99\textwidth]{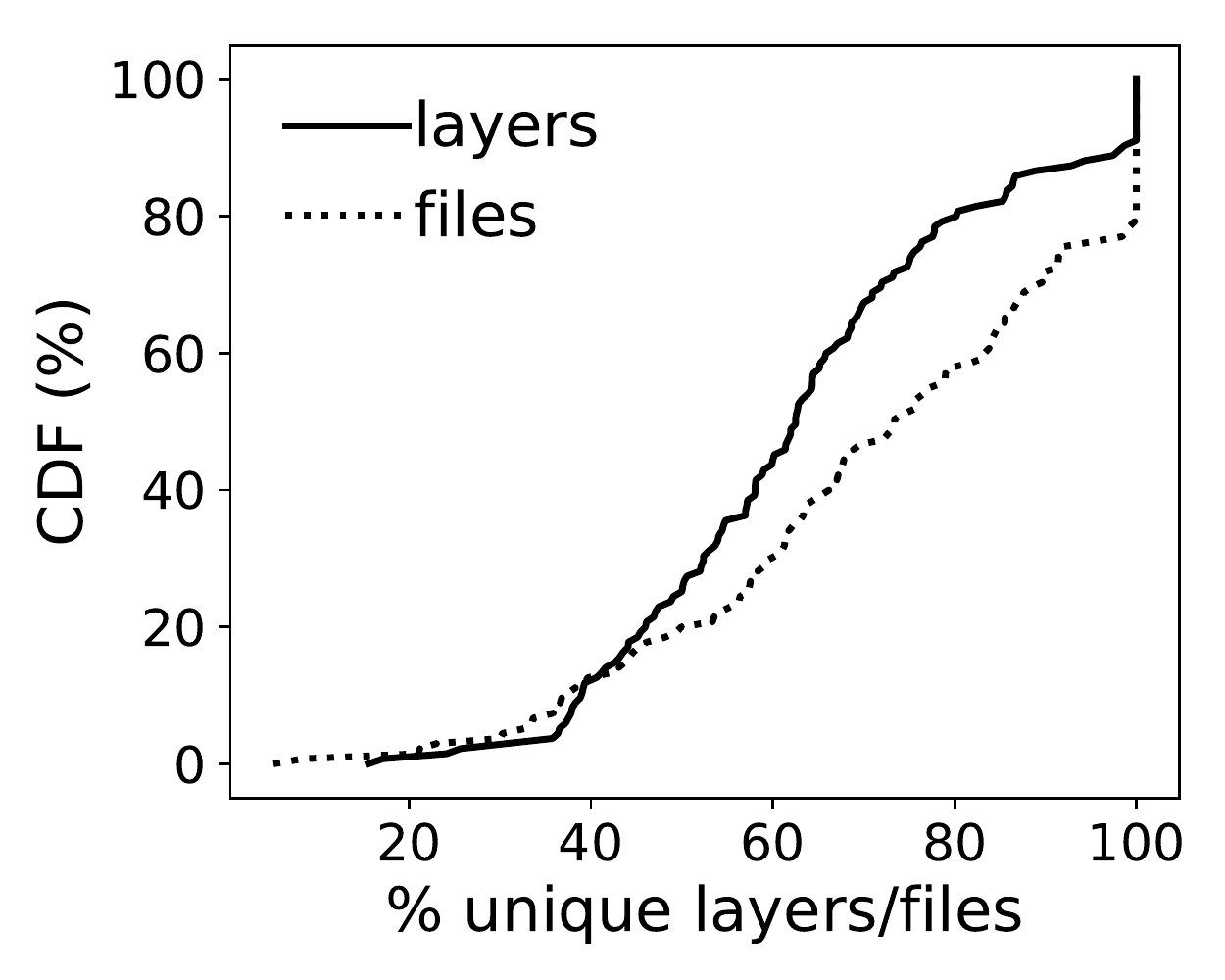}
  \vspace{-20pt}
  \caption{\small {\bf \% unique layers (files) among all\\ layers (files) in each official image repo.}} 
  \label{fig:download}
\end{minipage}
\hfill
\begin{minipage}[t]{0.315\textwidth}
	\centering   
  \includegraphics[width=0.99\textwidth]{analysis_x.pdf}
  \vspace{-20pt}
  \caption{\small {\bf \% files (layers) needed to analyze with official images as the knowledge base.}} 
  \label{fig:analysis}
\end{minipage}
\end{figure*}
}




\section{Challenges}
\label{sec:challenge}


\para{Image repositories are large in size.} Compared with code bases and apps, images often contain
orders of magnitude more files, because they encapsulate the entire systems environment
needed to run the target software, including OS, libraries (e.g., \texttt{libc}), 
runtime (e.g., JVM),
and tools and utilities.
As a result, images are typically orders of magnitude larger in size 
than code bases or even the entire code repositories.
Figure~\ref{fig:repo_size} shows the sizes of the 134 official image repositories on Docker Hub, 
compared with the size of corresponding code repositories of target software on GitHub. 
The image repositories are typically of sizes in gigabytes, with each
image being hundreds of megabytes, while the source code repositories are
in the range of tens to hundreds of megabytes.

As image repositories typically contain several tens of images with different tags, 
they could occupy up to tens of gigabytes in total 
(the sizes keep growing with new versions released).
Therefore, a statistically meaningful image dataset
(e.g., hundreds of repositories) would amount to the terabyte scale in total.

The challenges imposed by the excessive repository sizes 
are less at the storage level (as it can be mitigated by {\it stream}-based methods, \S\ref{sec:collection}),
but more for the bandwidth/time needed for fetching and analyzing images (downloading terabytes 
of data through the Internet). Images contain a large number of files, and thus need significant processing time
if all the files need to be iterated, though most of the files
may be irrelevant to the software engineering task. 


\begin{figure}[t]
\centering
  \includegraphics[width=0.31\textwidth]{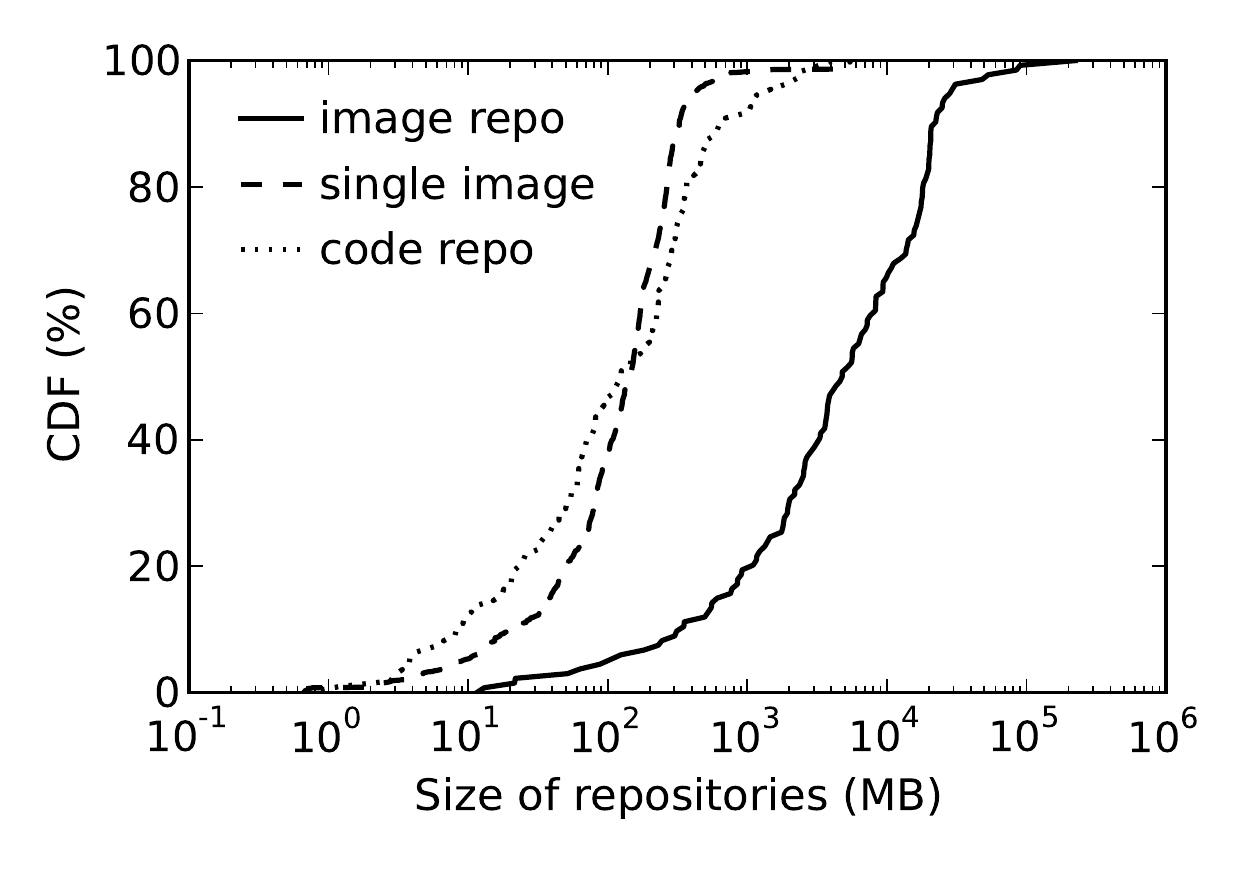}
  \vspace{-12pt}
  \caption{\small {\bf Size of image repositories versus code repositories.}} 
  \vspace{-8pt}
  \label{fig:repo_size}
\end{figure}

\para{Images are created with heterogeneous conventions.} 
The heterogeneity mainly comes from the underlying OS distributions
and configurations.
Even for the same version of software, images could be packed on top of 
different OS distributions (e.g.,
Debian vs. CentOS) which place binaries and configuration files at different
filesystem locations. Moreover, different images are equipped with different
tools (e.g., \texttt{apt} for Debian and \texttt{yum} for CentOS for managing packages and their dependencies). 
The pre-installed software components can also be heterogeneous: 
(1) certain packages (even those in \texttt{coreutils}) might not exist in all the images;
(2) different software variations can have incompatible requirements 
(e.g., different Unix \texttt{shell} variations have different syntax).

%% file: method.tex
\section{Mining Methods}
\label{sec:collection}

This section describes the methods for analyzing container image repositories, including
the process and techniques for addressing the challenges derived from the characteristics 
of images (\S\ref{sec:challenge}).

\para{Stream-based mining.}
Due to the large sizes of image repositories (cf. \S\ref{sec:challenge}), 
image repositories mining needs to adopt the \textit{stream}-based 
process 
if it cannot afford mirroring all the repositories locally.
A stream-based method extracts the target information continuously after 
images are loaded into memory/disks, and then removes these images to 
make space~\cite{shu:16}.
This can be done by either
{\it static} or {\it dynamic} method based on whether to \texttt{run} the images:

\begin{itemize}
  \item {\it static} methods analyze the 
\texttt{tar} archive of an image saved on local storage. As introduced
in \S\ref{sec:background}, an image is organized as a series of layers in the form of filesystem
\texttt{diff}s, which can be composed to create a unified filesystem hierarchy.
The files of interest can be extracted; 

\item {\it dynamic} methods first launch containers from the target images and then
  collect information of interest by invoking mining and analysis code inside the containers (which requires to
  \texttt{copy} the code into the container's filesystem and \texttt{copy} the analysis
    results from the container out to the host filesystem).  
  The code has the capability to invoke local commands and utilities available in the containers.
\end{itemize}

In comparison to dynamic methods, static methods are more lightweight (without the need to initialize/run containers);
they are also conceptually simpler as all the information is encoded in the files inside the \texttt{tar} archives
and can be analyzed through a uniform file-based processing framework.
On the other hand, dynamic methods can precisely capture runtime
information by directly executing commands in the containers.
However, this comes with the cost of complexity due to the diversity of containers (cf. \S\ref{sec:challenge}). 
For example, the mining code needs to consider different sets of pre-installed packages/utilities 
in terms of types, versions, and configurations.



\xcomment{
\para{Parallel processing.} Downloading and analyzing images in parallel with distributed
computing frameworks could significantly improve mining performance.
Note that image mining is I/O-intensive so the processing needs to 
avoid I/O contention. Although sophisticated frameworks like MapReduce can be leveraged, 
simple job distribution frameworks such as Redis Queue can be used to manage distributed mining when jobs are independent. 
}

\begin{figure}[t]
\centering
\begin{minipage}[t]{0.23\textwidth}
	\centering   
  \includegraphics[width=\textwidth]{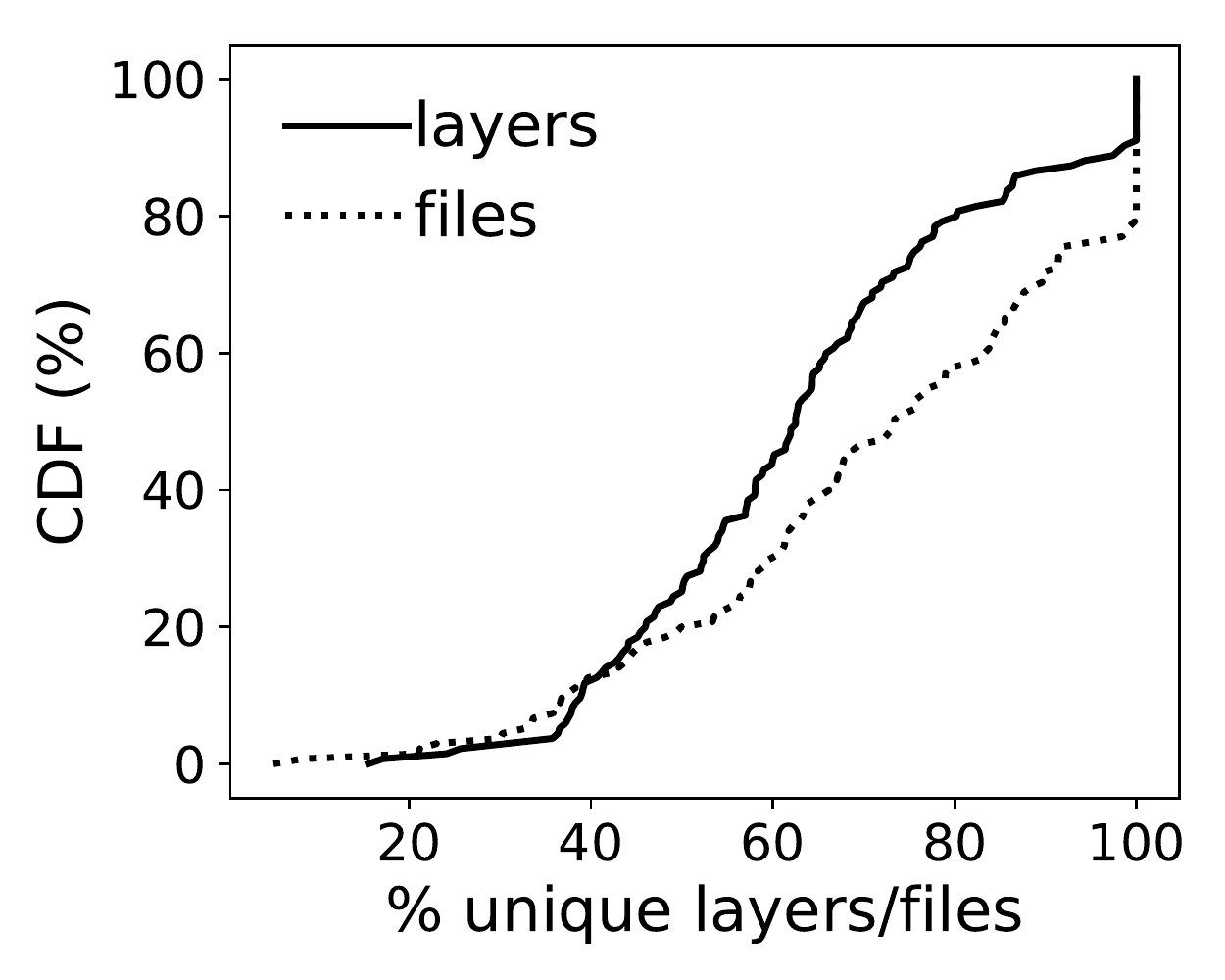}
  \vspace{-12pt}
  \caption{\small {\bf \% unique layers (files) among all layers (files) in each official image repository.}} 
  \label{fig:download}
\end{minipage}
\hfill
\begin{minipage}[t]{0.23\textwidth}
	\centering   
  \includegraphics[width=\textwidth]{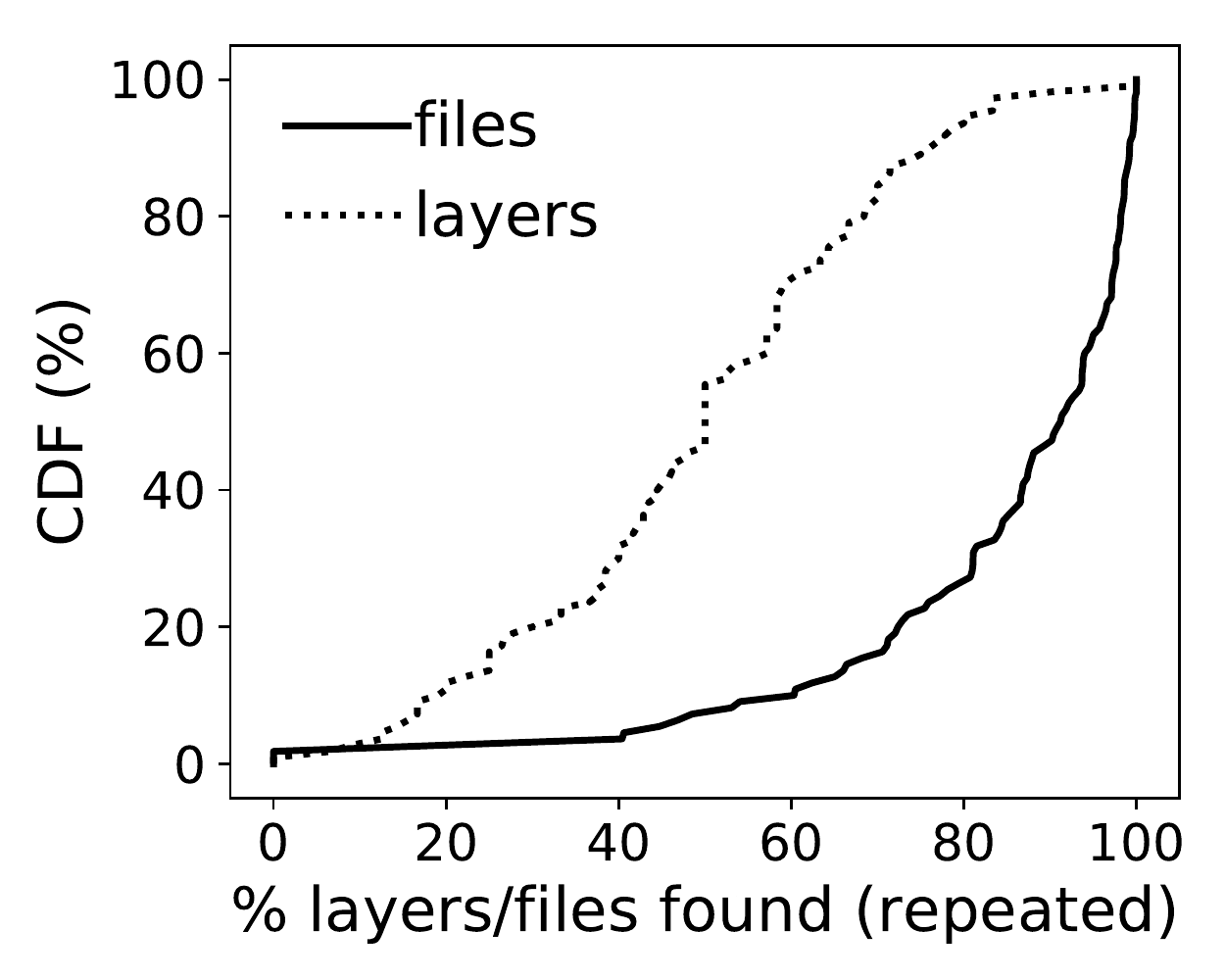}
  \vspace{-12pt}
  \caption{\small {\bf \% files (layers) needed to analyze with official images as the knowledge base.}} 
  \label{fig:analysis}
\end{minipage}
\end{figure}

\para{Downloading images with shared layers.} Images are organized in the granularity of layers (cf. \S\ref{sec:background}).
Each layer of an image is pulled down separately, and stored in the host machine.
If multiple images share the same layers (e.g., built upon the same OS image), these 
layers only need to be downloaded once. 
As a result, downloading images with shared layers in batches can save
significant storage and downloading overhead, compared with treating each image independently.
Typically, images from the same repositories share common layers and can be batched together, as they
likely share many base layers. 
Figure~\ref{fig:download} shows the percentage of unique layers across all the layers in each official repository
on Docker Hub (there are 143 official repositories)---batching the downloads can save 35+\% layers for 50+\% repositories.
A more sophisticated approach is to leverage the \texttt{FROM} instruction in Dockerfile 
that specifies the base image, from which the target images were built. 

\para{Layer-based analysis.} Similar to downloading, the image mining/analysis should
be designed and implemented based on layers. 
Layers that have been processed should be recorded to avoid repeated computing effort in a Dynamic Programming (DP) style.

To illustrate the efficacy of layer-based analysis, we pull the 143 official image repositories from
Docker Hub, and record the MD5 checksum of every file in each layer in a database (serving as the knowledge base). 
Then, we randomly sample 100 image repositories from Docker Hub and select the \texttt{latest} image in each repository. 
Figure~\ref{fig:analysis} shows the files with MD5 found in the knowledge base---on average, 
83.4\% of the files hit a small base of 143 official repositories,
even though the coverage of exact layers (based on LayerIDs) is much lower.
The main reason of such significant coverage of files
is that most files in an image come from the OS and libraries. As there are limited OS distributions
and library versions, a DP-style mining method can save significant (redundant) computing efforts.

\para{Selective mining.} Not every image in a repository is worth mining for a specific software engineering task.
For example, many images are for the same application binaries and configurations, but wrapped around different
OS distributions or libraries. If the information of interest lies in the application itself, only one of the images needs 
to be downloaded and analyzed.\footnote{Specifically, \texttt{Alpine} Linux is the OS distribution officially adopted by Docker since 2016, which is 
an order of magnitude smaller than \texttt{ubuntu} (the previous default).
Therefore, images based on \texttt{Alpine} are often the choice for downloading and analysis.}

\para{Leveraging Dockerfile.} A Dockerfile records
how an image is created (cf. \S\ref{sec:background}). The Dockerfile of an image can be fetched through Docker's REST API if available.
A lot of information of images can be collected and inferred by analyzing Dockerfiles, without the need to download and mine the images.
Unfortunately, as reported in~\cite{cito:17}, many Dockerfiles are not reproducible due to missing version pinning; moreover,
34\% of Dockerfiles were not able to build the images from a sample of 560 projects.

%% file: limitation.tex
\section{Limitations}

It should be noted that images only contain static information at the deployment time, 
but do not capture the dynamic information of running container instances.
For example, it is possible that the configuration settings are changed during the operation of the containers.
Therefore, the configurations stored in the images may not reflect the real usage in practice.
It will be beneficial to relate the data in container images to other data sources (e.g., runtime logs, performance counters, and workload characteristics), 
towards enabling richer and more insightful analysis.

Regarding software orchestration, though it can be understood better with Docker compose files,
there could be management operations outside the scope of containers and images driven 
by home-brewed scripts and procedures. 
Understanding the complete workflow and process of orchestration remains an open challenge.



%% file: related.tex
\section{Related Work}
\label{sec:related}

Prior studies on Docker images mostly focus on analyzing Dockerfiles as a special type of code~\cite{cito:17} 
and the security implications of adopting Docker images~\cite{shu:16,tak:17,combe:16}.
Differently, our focus is not about how they were created and
how secure to deploy them, 
but about the data and information that can be distilled from the images for the good and evil of software engineering research. 

Prior studies on mining software repositories mainly focus on 
source code repositories (including version-control systems and bug databases), 
archived communications, 
and app stores~\cite{xie:09,abate:15,hassan:08,harman:12,nachi:09,martin:17}. 
With the wide adoption of containerization techniques, container images have become emerging data which encode 
information unavailable in traditional software repositories. 
This paper advocates opportunities of mining container image repositories, as a special type of software repositories, to compliment prior work. 

Besides Docker images, virtual machine (VM) images are also available online, such as AMI (Amazon Machine Images) used for deploying VMs on Amazon EC2. On the other hand, AMIs do not have the same level of
popularity as Docker Hub. Moreover, AMIs do not have the notion of ``repositories'' but are traditional disk images 
which contain less semantic information. 



%% file: conclusion.tex
\section{Concluding Remarks}
In this paper, we advocate for mining container image repositories, as a special and emerging type of software repositories,
for understanding configurations and use cases of software systems. 
The motivation derives from the observation that few existing studies have paid attention to
container image repositories, or have explored the unique, rich data and information which is not available in traditional software repositories. 
To stimulate future research, 
we have discussed the opportunities of mining container image repositories, followed by the challenges and mining methods.
We hope that image repositories mining can fill the gap between in-house software development and the operations of software systems.


\section*{Acknowledgement}
We thank the anonymous reviewers of ICSE NIER for their valuable comments that help improve the presentation.
Darko Marinov's group is supported by National Science Foundation grants
CCF-1409423, CCF-1421503, CNS-1646305, and CNS-1740916.